\documentclass[aps,prb,twocolumn,superscriptaddress,showpacs]{revtex4}
\usepackage{bm}
\usepackage{amsmath}
\usepackage{amssymb}
\usepackage{graphicx}
\usepackage{color}
\begin{document}
\title{Umklapp-Assisted Electron Transport Oscillations in Metal
Superlattices}
\author{S. I. Kulinich}
\affiliation{B. Verkin Institute for Low Temperature Physics and
Engineering of the National Academy of Sciences of Ukraine, 47
Lenin Ave., Kharkov 61103, Ukraine} \affiliation{Department of
Physics, University of Gothenburg, SE-412 96 G{\" o}teborg,
Sweden}
\author{L. Y. Gorelik}\email{gorelik@chalmers.se}
\affiliation{Department of Applied Physics, Chalmers University of
Technology, SE-412 96 G\"{o}teborg, Sweden}
\author{S. A. Gredeskul}
\affiliation{Department of Physics, Ben Gurion University of the
Negev, POB 653 Beer-Sheva, 84105 Israel }
\author{R. I. Shekhter}
\affiliation{Department of Physics, University of Gothenburg,
SE-412 96 G{\" o}teborg, Sweden}
\author{M. Jonson}
\affiliation{Department of Physics, University of Gothenburg,
SE-412 96 G{\" o}teborg, Sweden} \affiliation{SUPA, Department of
Physics, Heriot-Watt University, Edinburgh EH14 4AS, Scotland, UK}
\affiliation{Department of Physics, Division of Quantum Phases and
Devices, Konkuk University, Seoul 143-701, Korea}

\date{\today}
\begin{abstract}
We consider a superlattice of parallel metal tunnel junctions with
a spatially non-homogeneous probability for electrons to tunnel.
In such structures tunneling can be accompanied by electron
scattering that conserves energy but not momentum. In the special
case of a tunneling probability that varies periodically with
period $a$ in the longitudinal direction, i.e., perpendicular to
the junctions, electron tunneling is accompanied by ``umklapp"
scattering, where the longitudinal momentum changes by a multiple
of $h/a$. We predict that as a result a sequence of
metal-insulator transitions can be induced by an external
electric- or magnetic field as the field strength is increased.
\end{abstract}

\maketitle To design novel electrical conductors in the form of
artificially structured materials remains one of the most
important tasks of nanoscience. This is because progress in this
type of ``quantum engineering" may lead to new and better
electronic devices. Multilayered systems are a widely used
material of this type, with semiconductor superlattices arguably
the most prominent example.\cite{01} Work on semiconductor
superlattices with spatially modulated properties on the
sub-micron scale started already 40 years ago, following the
pioneering work of Esaki and Tsu. \cite{02} The early focus on
semiconductors was natural, since the de-Broglie wavelength of
their conduction electrons is typically large enough to be
comparable to the period of then feasible superlattices.
Qualitatively new effects based on quantum interference phenomena
--- still mostly absent in metal superlattices\cite{Schuller}
--- could therefore be predicted and observed.\cite{01}

More recent developments have lead to engineered conductors such
as quantum dots, nano-wires, and other ``nanosolids", which could
be useful components in novel superlattice
architectures.\cite{RaphaelTsu} Here, we focus on nano-wires and
note that many are good metals with a high conductivity due to
ballistic electron transport. However, the de Broglie wavelength
of their electrons is typically much too small compared to the
modulation period $a$ of a nano-wire based superstructure for
quantum interference effects to occur. On the other hand, we will
show in this Letter that a prominent interference effect of a
different origin emerges in such structures under the realistic
assumption that $k_F a\gg1$, so that the quasi-classical
approximation is valid and hence the electron energy dispersion
can be linearized.

\begin{figure}
\centering
\includegraphics[width=0.85\columnwidth]{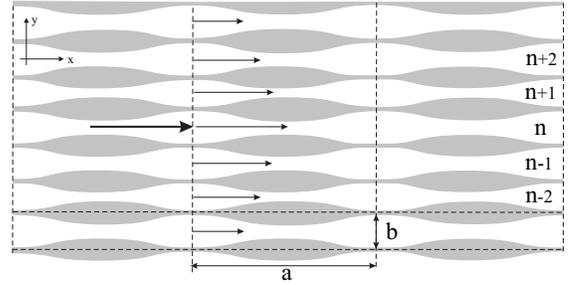}
\caption{Sketch of the considered $2D$ superlattice, comprising an
array (period $b$) of parallel nanowires coupled by inter-wire
electron tunneling with a periodically modulated tunneling
strength (period $a$).}\label{Fig1}
\end{figure}

The two-dimensional (2D) superlattice structure to be considered
is sketched in Fig.~\ref{Fig1}. It comprises a set of 1D wires
coupled by electron tunneling in such a way that the probability
for tunneling varies periodically along the direction of the
wires. For this structure we will show that when a magnetic field
is applied perpendicular to the 2D plane --- or when an electric
field is applied in the plane and perpendicular to the wires --- a
series of metal-insulator transitions occur with respect to the
inter-wire hopping transport of electrons as the strength of the
external field increases.

In order to understand how an increasing electric or magnetic
field can induce a series of metal-insulator transitions, it is
instructive to first consider a single, isolated wire, where
electrons move freely along the longitudinal direction and occupy
a 1D  band of ``longitudinal" energies. In the transverse
direction, they are confined to the wire and occupy a single,
discrete ``transverse" energy level. Now, disregard for a moment
the longitudinal motion and focus on the transverse electron
dynamics in a set of identical wires, aligned in parallel and each
with the same transverse level occupied. By switching on a
longitudinally uniform probability for electrons to tunnel to
adjacent wires, these previously degenerate energy levels will
form a band, which allows transverse motion between wires. Hence,
in the absence of an external field the superlattice is
effectively a two-dimensional (non-isotropic) metal.

If we now apply a external electric field $\overrightarrow{\cal
E}$ perpendicular to the wires, the transverse energy levels will
be shifted out of resonance so that band motion in the transverse
direction is prevented by Wannier-Stark localization of the
electron states.\cite{Wannier-Stark} However, if the tunneling
probability can be made to vary periodically along the wires, the
situation is qualitatively modified. This is because (i) the
longitudinal and transverse motion of the electrons can no longer
be separated, and (ii) the longitudinal momentum only has to be
conserved modulo $\hbar G$, where $G=2\pi/a$ and $a$ is the
modulation period. Although the total energy is still conserved
when electrons tunnel between wires, energy can now be shifted
from the longitudinal to the transverse motion in ``umklapp
processes" that involve discrete changes of longitudinal momentum.
It follows that the transverse energy-level shifts can be
compensated and band motion restored for a discrete set of
electric-field values.

The effect of umklapp-assisted resonant tunneling is controlled by
the dimensionless parameter ${\cal \phi}= \Phi/\Phi_0$, where
$\Phi=H_{\rm eff}\,ab$ is the flux of an ``effective" magnetic
field $H_{\rm eff}=(c/v_F)\vert \overrightarrow{\cal E}\vert$
through a superlattice ``unit cell" of area $ab$ (see
Fig.~\ref{Fig1}) and $\Phi_0=hc/e$ is the magnetic flux quantum.
The result is the same if instead a magnetic field $H$ is applied,
except that now $\Phi=Hab$. Resonant tunneling occurs when ${\cal
\phi}=p/q$ is a rational number ($p$ and $q$ are integers) but, as
we will discuss below, band motion is only possible when ${\cal
\phi}$ is an integer (i.e., for $q=1$).

\begin{figure}
\centering
\includegraphics[width=0.85\columnwidth]{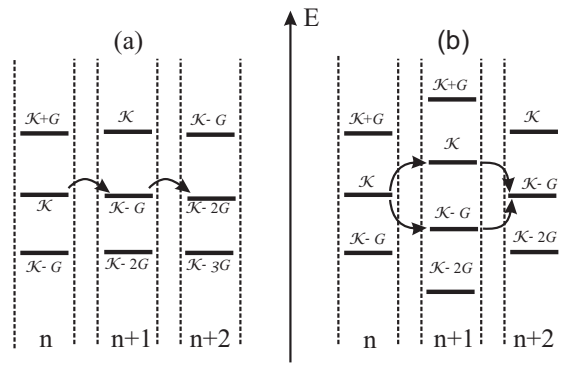}
\vspace*{0.5cm} \caption{Transverse energy levels involved in
resonant inter-wire tunneling (thick bars) in three neighboring
wires ($n$, $n+1$, $n+2$). The total electron energy is the sum of
a longitudinal part $\propto {\cal K}$ (linear spectrum assumed)
and a transverse part, which (for fixed ${\cal K}$) is shifted
from one wire to the next by an amount proportional to the
dimensionless flux $\phi$ of an external electric or magnetic
field. Since ${\cal K}$ in the periodically modulated superlattice
system of Fig. \ref{Fig1} is only conserved modulo $G$, energy can
be shifted between the longitudinal and transverse parts in a
tunneling event that conserves the total energy. This is why a
sequence of transverse levels is shown for each wire. For integer
values of the flux parameter, as in (a) where $\phi=1$, resonant
transmission (arrows) between states on adjacent wires are
energetically allowed. In (b), where ${\cal\phi}=1/2$, resonant
transitions occur between states on every second wire. The two
paths through virtual states on the intermediate wire shown
(arrows) contribute with equal amplitude but opposite signs to to
the total transition amplitude, which when all paths are
considered turns out to be zero due to destructive interference.}
\label{Fig2}
\end{figure}

In Fig.~\ref{Fig2} we illustrate the effect of umklapp-assisted
resonant tunneling when ${\cal \phi}=1$ (panel a) and
${\cal\phi}=1/2$ (panel b). While in case (a) the resonance
condition is fulfilled for neighboring wires, in case (b) every
second wire is in resonance and resonant tunneling occurs via
virtual states on the intermediate wire. In this case
contributions to the total tunneling amplitude from a number of
different paths through various virtual  states have to be summed
up. The contributions from the two paths identified by arrows in
Fig.~\ref{Fig2}(b) have equal magnitude but different signs (since
the two virtual levels have mirror symmetry with respect to the
resonant levels). By generalizing this argument one finds that
tunneling between the resonant levels becomes completely
suppressed by destructive interference in this case (if direct
hopping between next-nearest neighbors is neglected). In the
general case of a rational flux, ${\cal\phi}=p/q$, the transverse
energies are resonant for wires separated by a distance $\Delta
y=qb$, where $b$ is the superlattice period. Although it would be
quite difficult to explicitly consider the destructive
interference between all the possible paths for large $q$, we are
nevertheless able to prove below that resonant hopping is
completely suppressed for any (non-integer) rational value of the
parameter $\cal\phi$.

We consider an infinite superlattice structure (see
Fig.~\ref{Fig1}) subject to a constant electric field
$\overrightarrow{\cal E}=-{\cal E}{\hat y}$. In the quasiclassical
limit ($k_Fa\gg 1$) large momentum transfers ($\Delta p\sim p_F)$
may be neglected, which justifies a linearization of the energy
dispersion related to the longitudinal motion (in each wire).
Also, the Hamiltonian may be split into separate parts for left-
and right-moving electrons, which can be treated independently.
Hence, the spectrum $E$ (measured from the Fermi level) and the
stationary wave functions for the problem at hand can be found by
solving the Schr\"odinger equation
\begin{eqnarray}\label{04}
-E_0\left(i\frac{\partial}{\partial x} + 2\pi n \phi  \right)
\varphi_n(x)+\hspace*{3.5cm} \\
v(x)\left[\varphi_{n+1}(x)+\varphi_{n-1}(x)\right]=E \varphi_n(x)\,.
\nonumber
\end{eqnarray}
Here $x$ is the dimensionless coordinate along the wire(s),
$v(x)=v(x+1)$ is the periodic potential responsible for electron
transitions between the wires, and $E_0=\hbar v_F/a$. Equation
(\ref{04}) is unchanged if one instead considers a superlattice
subject to a constant and not too strong (see below) magnetic
field $\overrightarrow{H}=H{\hat z}$. In both cases $\cal \phi$ is
the dimensionless flux of the relevant external field --- the
magnetic field $H$ or the effective magnetic field $H_{\rm
eff}=(c/v_F){\cal E}$ --- through a unit cell of the superlattice
structure.

We will now proceed by solving Eq.~(\ref{04}) exactly for an
arbitrary potential $v(x)$. The first step is to note that
according to Bloch's theorem $\varphi_n(x)=\exp{(i{\cal K}x)}
u_n(x)$, where ${\cal K}$ is the (dimensionless) quasimomentum,
$-\pi<{\cal K}<\pi$, and $u_n(x)=u_n(x+1)$ is a periodic function.
It is convenient to define the auxiliary function $u(n)\equiv
u_n(x=-1/2)$. This function obeys the equation,
\begin{equation}\label{9}
u(p)=e^{i(\varepsilon-{\cal K}+2\pi{\cal\phi} p)}\sum_n
e^{-i \theta(p-n)}J_{p-n}(A)u(n),
\end{equation}
where $J_n(x)$ is a Bessel function, $\varepsilon=E/E_0$, while $A$
and $\theta$ are defined through the relations
\begin{equation}\label{10}
\frac{2}{E_0}\int_{-1/2}^{1/2}dy v(y)e^{ 2\pi i{\cal\phi}
y}\equiv Ae^{i \xi},\,\,\theta=\pi/2+\pi{\cal\phi}+\xi.
\end{equation}

Equation (\ref{9}) determines the energy spectrum and the wave
functions in our problem. The structure of the spectrum  strongly
depends on the nature of the number $\cal \phi$. If $\cal \phi$
has a  {\it non-integer} value, the eigenenergies and eigenstates
are labeled by the three quantum numbers ${\cal K},\, m,$ and $r$;
the quasi-momentum $\cal K$ and the (integer) band index $m$ refer
to the longitudinal motion along the $x$-axis while the integer
$r$ is related to the transverse motion in the $y$-direction. The
dispersion law reads
\begin{equation}
\label{011} E_{m,r}({\cal K})=E_0\left({\cal K}+2\pi
m-2\pi{\cal\phi} r\right),\,\,m,r=0, \pm1,...
\end{equation}

The energy-level distribution depends crucially on whether or not
the noninteger $\cal \phi$ is a rational number. If it is a
rational number, ${\cal \phi}=p/q$, one notes that $E_{m,r}({\cal
K})=E_{m+Mp,\,r+Mq}({\cal K})$ for a given quasi-momentum $\cal K$
and any integer $M$. This results (for each $\cal K$) in a set of
infinitely degenerate, equidistant energy levels. If $\cal \phi$
is an irrational number, on the other hand, the energy levels are
homogeneously distributed forming a discrete spectrum that is said
to be everywhere dense.

The eigenfunctions can be found from Eq.~(\ref{9}) rewritten as
\begin{equation}\label{12}
u_{m,r}(n)=e^{-i \xi (n-r)}J_{n-r}(\gamma),\,
\gamma=\frac{A}{2\sin\pi\cal\phi},
\end{equation}
and have the same form for both rational and irrational values of
$\cal\phi$. Therefore,  for any non-integer $\cal \phi$, all
states are localized near a particular wire, $r$, within a
localization radius $R_{\rm loc}$ defined as
\begin{equation}\label{14}
\left(\frac{R_{\text{loc}}}{b}\right)^2 \equiv \sum_n n^2
J_{n-r}^2(\gamma)-\left[\sum_n n
J_{n-r}^2(\gamma)\right]^2=\frac{\gamma^2}{2}.
\end{equation}
It is remarkable that this result holds even when the flux
parameter $\cal\phi$ is a (non-integer) rational number, since in
this case our superlattice structure has translational symmetry in
the $y$-direction. Accordingly, bands of electron states with
infinite extension in this direction should form (see, e.g.,
Ref.~\onlinecite{5}). However, even though the transverse energy
levels of the parallel wires periodically are in resonance, this
does not happen. The reason is a fully destructive interference
between the probability amplitudes for resonant hopping along
different paths, as illustrated for the special case of
${\cal\phi}=1/2$ in Fig.~\ref{Fig2}(b).\cite{03}

In case $\cal \phi$ is an {\it integer}, not only does the
external field become effectively periodic in the $y$-direction
but a band of extended states also form [$R_{\rm loc}\to\infty$,
according to Eqs.~(\ref{12}) and (\ref{14})]. Therefore, in
addition to the two ``longitudinal" quantum numbers ${\cal K},m$ a
continuous quasi-momentum ${\tilde{\cal K}}$, where
$-\pi<\tilde{\cal K}<\pi$, must be used to label the transverse
motion along the $y$-axis. The dispersion law, which is found from
Eq.~(\ref{9}), reads
\begin{equation}\label{16}
E_m( {\cal K},\,\tilde {\cal K})=E_0 \left({\cal K}+2\pi m+A\sin
(\tilde {\cal K}+\theta)\right).
\end{equation}
Thus, for each pair (${\cal K}, m$) of eigenvalues related to the
longitudinal motion, the transverse energies spread into a band of
width $\delta E=2AE_0$. The corresponding eigenstates are of the
plane wave type, $u_{{\tilde{\cal K}}, m}(n)=\exp(-i\tilde {\cal
K}n)$ and, as a consequence, are delocalized in both the
longitudinal and transverse directions.

The detailed features of the energy spectrum influence various
physical quantities in essential ways. Here, we consider the
linear response of the system to a weak ac external electrical
field $\overrightarrow{{\cal E}}(t)={\cal E}_0\cos \omega t
\,{\hat y}$. The interaction with the field adds a perturbation
term ${\cal H}_{\rm int}(t)={\cal H}^{(\text{int})} \cos\omega t$
to the Hamiltonian, where
\begin{equation}\label{15}
{\cal H}^{(\text{int})}= eb{\cal E}_0\sum_nn\int dx\Phi_n
^\dag(x)\Phi_n(x)\,,
\end{equation}
and $\Phi_n^\dag (x)$ [$\Phi_n(x)]$ is a field operator that
creates [destroys] an electron at point $x$ in the $n$-th wire
and obeys the standard anti-commutation relations.

Absorption of the ac electric field is proportional to the real
part of the conductivity  $\sigma(\omega)$, which in linear
response theory has the form
\begin{equation}\label{17}
\sigma(\omega) =\frac{1}{{\cal E}_0L}\sum_{\alpha,\beta}
I_{\alpha,\beta}{\cal H}^{(\text{int})}_{\beta,\alpha}
\frac{f(E_\alpha)-f(E_\beta)}{E_\alpha-E_\beta-\hbar(\omega-
i\nu)}.
\end{equation}
Here $\nu$ is a phenomenological relaxation rate, $L$ is the
sample length in the longitudinal direction, $f(E)$ is the
Fermi-Dirac distribution function, and $I_{\alpha,\beta}$ is the
matrix element of the current operator
\begin{equation}\label{18}
\hat I=\frac{i e}{\hbar N}\sum_n\int dx v(x)\left[\Phi_{n+1}
^\dag(x)\Phi_n(x)-H.c.\right],
\end{equation}
where $N$ is a total number of wires in superlattice.

Standard calculations lead to an electrical conductivity of the
form
\begin{eqnarray}\label{22}
\sigma(\omega)= \frac{i e^2}{h}\frac{ab}{(\hbar
v_F)^2}\sum_n\vert v_n\vert^2\times\qquad\qquad\\
\times\left[\frac{1}{ 2\pi(n-{\cal\phi})-t_0(\omega-i \nu)}
-\frac{1}{ 2\pi(n-{\cal\phi})+ t_0(\omega-i \nu)}\right],
\nonumber
\end{eqnarray}
where $t_0=a/v_F$ and $v_n$ is a Fourier component of the
potential $v(x)$. In the low frequency limit, $\omega\ll\nu$, the
electrical conductivity can be approximated as
\begin{equation}\label{23}
\sigma(0)=G_0\frac{ab}{(\hbar v_F)^2}\sum_n\vert v_n\vert^2
\frac{t_0 \nu}{[2\pi(n-{\cal \phi})]^2+(t_0 \nu)^2},
\end{equation}
where $G_0=2e^2/h$ is the conductance quantum.
\begin{figure} \vspace*{-0.5cm}
\centering
\includegraphics[width=0.75\columnwidth]{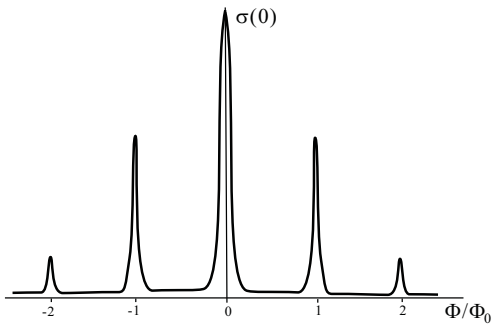}
\vspace*{-1cm}
 \caption{Schematic behavior of the low-frequency
transverse conductivity, $\sigma(0)$, as a function of magnetic
field flux $\Phi$ through a unit cell of the superlattice
structure shown in Fig.~\ref{Fig1};  $\Phi_0=hc/e$ is the flux
quantum.} \label{Fig3}
\end{figure}
It follows that in this limit the conductivity plotted as a
function of the number of flux quanta per unit cell in the
superlattice, $\cal \phi$, has a set of peaks corresponding to
integer numbers, $\phi=0, \pm 1, \pm 2, \ldots$ (see
Fig.~\ref{Fig3}). The scale of the fluctuations are determined by
the ratio between successive maxima [$\sigma_{l}(0)$, $\phi=l$]
and minima [$\sigma_{l+1/2}(0)$, $\phi=l+1/2$] of the
conductivity,
\begin{equation}\label{24}
\frac{\sigma_{l}(0)}{\sigma_{l+1/2}(0)}\sim \frac{1}{(t_0\nu)^2};
\,\,\, \sigma_{l+1/2}(0) \sim G_0\,t_0\nu\left(R_{\rm loc}/b\right)^2
\end{equation}

\noindent where $R_{\text{loc}}$ is the localization radius,
Eq.~(\ref{14}). Hence, if $t_0\nu\ll 1$, the field dependence of
the absorption as well as the conductivity has a pronounced peak
structure as schematically shown for the conductivity in
Fig.~\ref{Fig3}.

In our analysis we have used the Schr\"odinger equation
(\ref{04}), which has a linear energy spectrum. For this
approximation to be valid the longitudinal momentum fluctuations
associated with umklapp-assisted resonant transmission of
electrons between adjacent wires must be small on the scale of the
Fermi momentum $p_F$. The total momentum change due to such
processes is restricted by the relaxation time $\tau_r\sim 1/\nu$,
which corresponds to coherent tunneling through $N_\nu\sim v_t
\tau_r/b$ wires, where $v_t\sim AbE_0/\hbar$ is the electron
velocity in the transverse direction. Reasonable estimates for the
``tunneling" parameter, $A\sim 10^{-3}$, and for the superlattice
periods, $a,\,b\sim$~10~$\mu$m, give an upper limit of
$\tau_r\sim$~10~ns for a relaxation time consistent with an
approximately linear spectrum. On the other hand, the criterion
that the electron motion is ballistic gives a lower limit for the
relaxation time of $\tau_r\sim b/v_t\sim$~0.1~ns.

An additional requirement for the Schr\"odinger equation
(\ref{04}) to be valid must be fulfilled if an external magnetic
field is applied to the system. This is because the cyclotron
motion of the electrons can be neglected (and a term quadratic in
$H$ dropped from Eq.~(\ref{04})) only if
\begin{equation}\label{37}
H\leq\frac{E_F}{ E_0}\,\frac{\Phi_0}{2\pi ab N_\nu}.
\end{equation}
For a relaxation time $\tau_r\sim$~1~ns this restriction
corresponds to  $H\leq$~6~T.

The single-particle approach used in our analysis neglects the
Coulomb interaction between electrons. In principle, however, a
Coulomb blockade of inter-wire tunneling might prevent the
formation of extended electron states in the transverse
direction.\cite{7} To avoid such a blockade the electrostatic
charging energy of the wire, $E_{el}\sim e^2/L$, should be smaller
than than width, $\delta E\sim AE_0$, of the transverse energy
band. Using our estimates for the relevant parameters, this
corresponds to a lower limit of order 100~$\mu$m for the length
$L$ of the nano-wires.

Finally we estimate the required strengths
of the external electric- and magnetic
fields by noting that a dimensionless flux of ${\cal\phi}=1$
corresponds to ${\cal E}\sim$~0.05~V/cm or $H\sim$~0.4~T.
Such field strengths can easily be applied in an experiment.

In conclusion, we have shown that the transport properties of a
superlattice comprising a set of parallel metallic nano-wires
coupled by tunneling in such a way that the tunneling probability
varies periodically along the wires drastically differ from the
predictions of linear transport theory. In particular, an
electric-field induced sequence of metal-insulator transitions
gives rise to a highly non-linear current-voltage characteristics,
while the sensitivity to a magnetic field leads to large-magnitude
oscillations of the magneto-conductance. Importantly, these
phenomena are manifest in comparatively weak external fields.
Hence, in relatively weak fields interference phenomena give rise
to pronounced mesoscopic features in the transport properties of
the studied metallic superlattices. Such superlattice structures
could, e.g., be realized using arrays of nanowires similar to
those that have been proposed for memory\cite{memory_array} and
mechanical single-electron transistor\cite{nanomech_shuttle}
applications.

Financial support from the Swedish Research Council (VR), the
European Commission (Grant No. FP7-ICT-2007-C; Project No. 225955
STELE), and the Korean WCU program funded by MEST/KOSEF
(R31-2008-000-10057-0) is gratefully acknowledged. SAG and SIK
acknowledge the hospitality of the Physics Department, University
of Gothenburg.


\begin{thebibliography}{99}
\bibitem{03}
Band motion in the $y$-direction can occur, however, if direct
tunneling between non-neigboring wires is allowed.
\end{thebibliography}
\end{document}